\documentclass[numreferences]{kluwer}
\usepackage{color}
\usepackage{graphicx}
\let\oldverbatim\verbatim
\renewcommand{\verbatim}{\expandafter\small\oldverbatim}
\def\spose#1{\hbox to 0pt{#1\hss}}
\def\lta{\mathrel{\spose{\lower 3pt\hbox{$\mathchar"218$}}
     \raise 2.0pt\hbox{$\mathchar"13C$}}}
\def\gta{\mathrel{\spose{\lower 3pt\hbox{$\mathchar"218$}}
     \raise 2.0pt\hbox{$\mathchar"13E$}}}

\def\eg{{\it e.g.,\,}}
\def\ie{{\it i.e.,\,}}
\def\etal{{\it et al.\,}}
 
\def\etc{{\it etc.\,}}

\def\be{\begin{equation}}
\def\ee{\end{equation}}
\def\bea{\begin{eqnarray}}
\def\eea{\end{eqnarray}}

\def\tauH{\tau_0}

\def\cobedmr{{\scshape cobe--dmr\,}}
\def\chils{{\chi_{{}_\mathrm{LS}}}}

\def\cl {{\cal C}_\ell}
\def\today{\ifcase\month\or
 January\or February\or March\or April\or May\or June\or
 July\or August\or September\or October\or November\or December\fi
 \space\number\day, \number\year}

\def\plotone#1{\centering
\includegraphics[width=0.8\columnwidth]{#1}}

\def\plottwoside#1#2{\centering
  \includegraphics[width=.48\columnwidth]{#1}\hfil
  \includegraphics[width=.48\columnwidth]{#2}}

\runningtitle{Probing beyond the cosmic horizon}
\runningauthor{Tarun Souradeep}

\begin{document}

\begin{article}

\begin{opening}

\title{ Probing Beyond the Cosmic Horizon}



\author{Tarun \surname{Souradeep}}

\institute{Department of Physics, Kansas State University, 
		Manhattan, KS 66506, U.S.A.}




\begin{ao}
116 Cardwell Hall, Kansas State University, Manhattan, KS 66506-2601, U.S.A.
\end{ao}


\begin{abstract}
There are reasons to believe that the region of uniform curvature that
we observe within our horizon is perhaps a tiny patch, much smaller
than the length scales of inhomogeneity and global connectivity of an
extremely complicated manifold. However, the recent supernova (SN)
results suggest that the horizon scale could be comparable to or even
much larger than curvature radius. Non trivial global structure too
tends to be of the order of the local curvature scale. Probing
(slightly) beyond the cosmic horizon can potentially reveal nontrivial
global structure lurking around or just beyond the horizon scale. The
cosmic microwave background (CMB) anisotropy is a sensitive probe of
the universe on length scales up to and somewhats beyond the horizon
scale and is perhaps poised to detect or put interesting limits on non
trivial features the global structure. This point is exemplified by
high CMB anisotropy signal of the Elliptical topology in a large
fraction of the revised parameter space of the closed FRW universe
(spherical geometry) currently preferred by SN observations.
\end{abstract}

\begin{motto}[prose]
\begin{flushright}
\itshape
How can I\\
even for an instant understand the beginning, the end,\\
the meaning, the theory -- of something outside of which \\
I can never go? Only this I know -- \\
that this thing is beautiful, great, terrifying,\\
various, unknowable, my mind's ravisher...\\
-- {\em Rabindranath Tagore}, (1901), \\
translated from Bengali\cite{tagore_quote}
\end{flushright}

\end{motto}


\abbreviations{ CMB -- cosmic microwave background; CH -- compact
hyperbolic; FRW -- Friedmann-Robertson-Walker; SLS--sphere of last
scattering; \cobedmr -- Cosmic Background Explorer-Differential
Microwave Radiometer; SN -- (high redshift) supernova }

\classification{JEL codes}{}
\end{opening}

\section{Introduction}

I am delighted to contribute to the festschrift honouring an
illustrious scientific career such as that of Jayant Narlikar and to
pay my tribute to a great teacher.  It is indeed an amazing and
fortunate coincidence that this festschrift comes at a time when my
research interests have wandered back into the interests of my early
graduate student days under Jayant's supervision. I owe my continued
interest in cosmic topology to his whole hearted encouragement and
tolerance towards rather crazy and even incorrect ideas at the start
of my research career.

One of the first scientific papers on cosmology that I read during
that phase was Jayant's paper (with Seshadri) addressing the
observability of Elliptical universes through the `counter-images' of
cosmologically distant light sources (say, quasars or very high
redshift galaxies)~\cite{ses_nar85}. In discussing the prospects of
probing beyond the horizon, it seems appropriate for the occasion to
present results from an ongoing work of mine that addresses the same
question using a different probe and, more importantly, in the light
of some very exciting recent observations.

\begin{figure}
\plotone{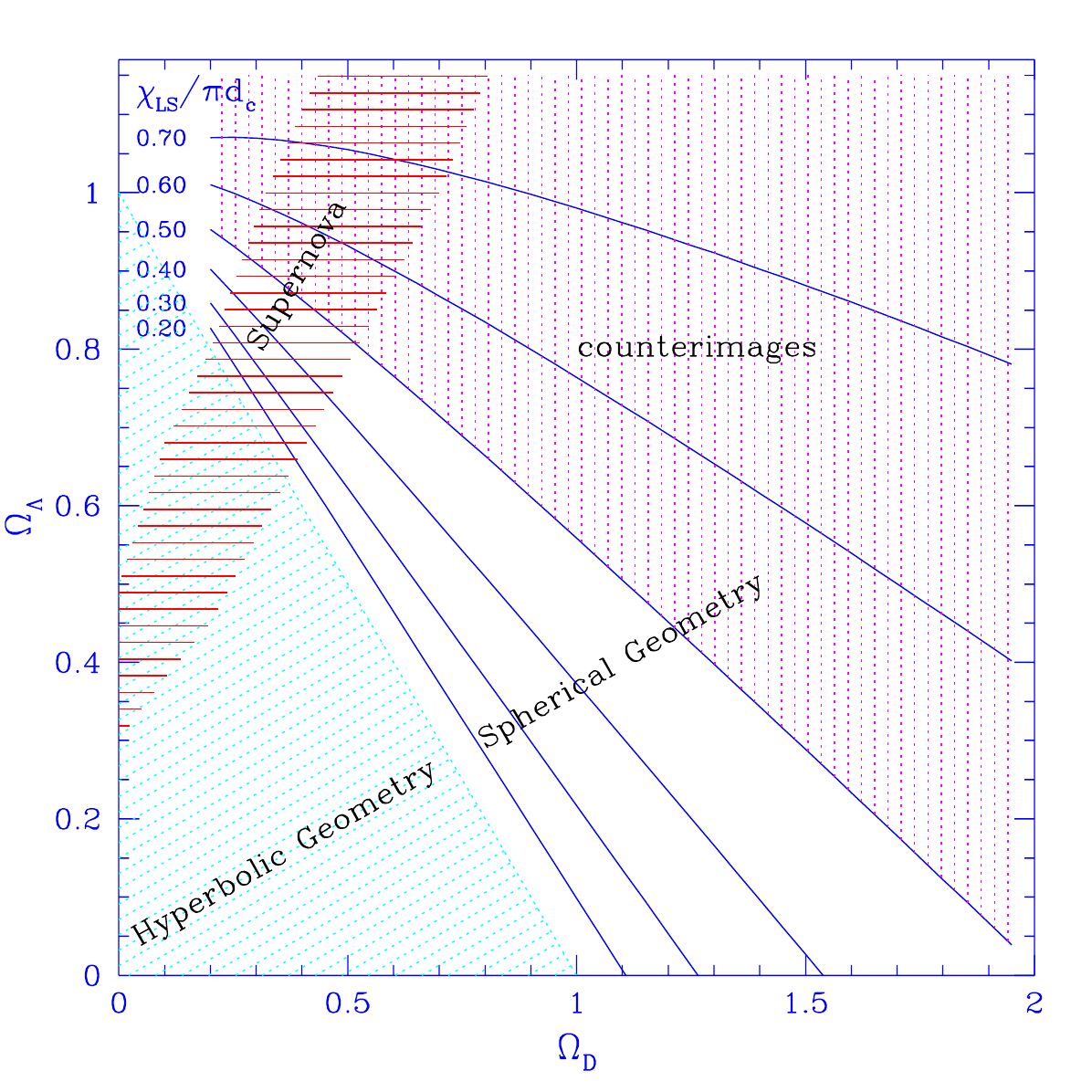}
\caption{The contours of constant ratio between the horizon radius and
curvature radius, $\tauH/d_c$ are shown on the
$\Omega_D-\Omega_\Lambda$ plane for models with spherical spatial
section. The models with hyperbolic geometry in the hatched lower left
corner are not considered here. The dividing line represents Euclidean
models ($d_c\to\infty\Rightarrow\tauH/d_c \to 0$). In the absence of a
cosmological constant ($\Omega_\Lambda=0$), counter-images lie within
the horizon (\ie satisfies $\tauH/d_c > \pi/2$) when $\Omega_D > 2$,
whereas, in the presence of a $\Lambda$ term the condition can be
satisfied for all $\Omega_D$ (vertically hatched region). Recent high
redshift supernovae measurements point to the presence of a
cosmologically significant $\Lambda$ term; the horizontally hatched
region in the parameter space is a rough depiction of the range of
models preferred by the Supernova results (part of the $68\%$
likelihood region that lies within the range of the plot; the peak
lies outside further into the Spherical and high $\tauH/d_c$ region at
around $(\Omega_D,\Omega_\lambda)\approx (0.75, 1.4)$). For most of
the preferred spherical models, the possibility of Elliptical topology
can be easily verified or refuted using the CMB anisotropy data}
\label{fig:omd_omv_plane} 
\end{figure}

\begin{figure}
\plotone{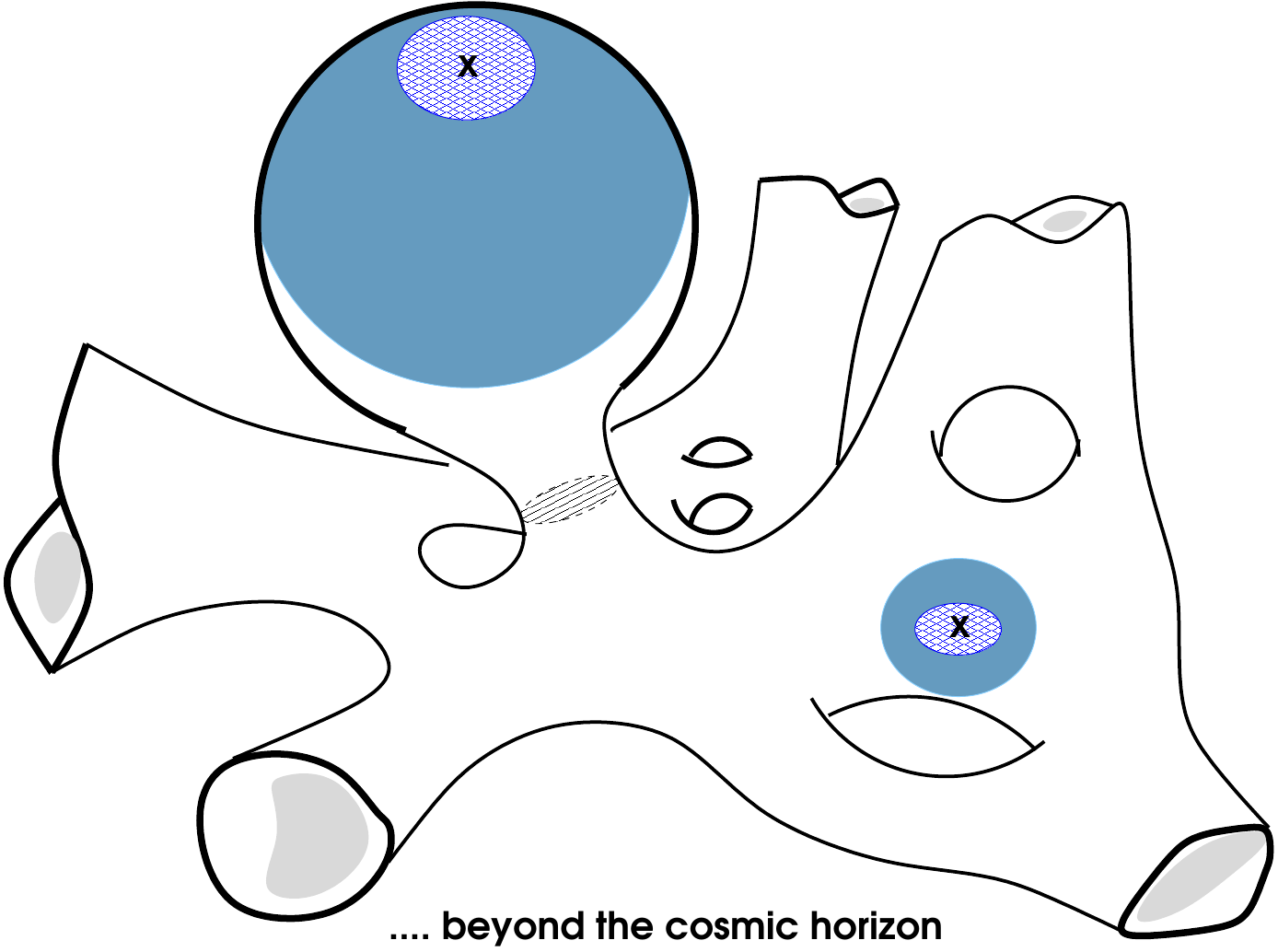}
\caption{ A cartoon of the possibly rich global structure of universe
on ultra-large scales is depicted. The `X' marks denote two observers
in regions with locally spherical and hyperbolic geometry,
respectively. In cosmological settings such that $\tauH/d_c << 1$, \ie
observable volume shown by the cross hatched discs around the observer
is small), the observer will never detect the rich global structure
and be satisfied with a standard ``open'' or `closed' FRW model of the
universe. However, in a cosmology such $\tauH/d_c$ is of order unity
(the larger shaded discs around each `X'), certain signatures of
global structure may be detectable. Note region within the observable
volume need not encompass any significant curvature variation, or be
multiply connected. The non-trivial global structure could reveal
itself in the CMB anisotropy through the modification of the spectrum
of fluctuations; \eg the narrow throat in the spherical case (boundary
of the hatched ellipse) and the `toroidal' regions in the hyperbolic
case.}
\label{fig:bub_univ} 
\end{figure}

The recent results from the high redshift supernova searches are
pointing to the presence of a cosmological constant,
$\Lambda$\cite{SNres}.  A non-zero $\Lambda$ term \footnote{ It
implies, more generally, some exotic form of matter that accelerates
the expansion of the universe. In this article, all my comments
regarding the possible $\Lambda$ term also applies to these other
exotic possibilities.} allows for values of the `horizon' size
relative to the curvature scale, $\tauH/d_c$ of order unity in
spherical models for acceptable values of $\Omega_D$ (see
figure~\ref{fig:omd_omv_plane}). {\em An important and exciting
ramification of this result is that nontrivial features in the global
spatial structure of universe may be within the grasp of
observations.}  The argument has two parts. First, the length scale in
the simplest forms of multiple connectivity (nontrivial topology) or
breakdown of homogeneity is of the order of the curvature radius,
$d_c$. Examples demonstrating this fact are plenty: the connecting
stalk in a bubble universe (see figure~\ref{fig:bub_univ}) can be at
most at a distance $\pi d_c$ from any observer on the smooth spherical
region; compact hyperbolic spaces have diameters~\footnote{The
diameter of a manifold is the maximum of distances between all
possible pairs of points.} of the order of $d_c$; the diameter of an
elliptical universe is $\pi d_c/2$, \etc. The second part is the fact
that the observed CMB anisotropy probes the spatial structure of the
universe up to scales comparable to and somewhats beyond the
horizon. In this article, I illustrate this point with the example of
Elliptical universe models.

There are theoretical motivations to believe that the spatial section
of the universe is indeed a complex manifold such as shown in
figure~\ref{fig:bub_univ}. That is not tractable in full generality.
A more tractable and well defined problem would be to restrict
attention to homogeneous (uniform) curvature manifolds with non
trivial topology~\cite{ntt}. The CMB anisotropy and the constraints
from data have been studied in many multiply connected Euclidean and
Hyperbolic universe~\cite{rev_ref}. Spherical geometry did not get
attention possibly because in a dust filled universe, $\tauH/d_c \ll
1$ for reasonable values of its cosmological density.  However, in the
presence of a non zero $\Lambda$, that is no longer true and the
observability of an Elliptical universe is an interesting question.

\section{CMB anisotropy in a Elliptical universe}
\label{ellip}

The realization that the universe with the same local geometry has
many different choices of global topology is as old as modern
cosmology -- De Sitter was quick to point out that Einstein's closed
static universe model with spherical geometry ${\cal S}^3$ could
equally well correspond to a multiply connected {\em Elliptical}
universe model where the antipodal points of ${\cal S}^3$ are
topologically identified~\cite{rev_ref}.

In order to keep within the suggested size for the article, I restrict
my text to points of direct relevance to the problem at hand. The
reader is directed to \cite{ses_nar85} for a good description
of Elliptical universe; excellent recent reviews~\cite{rev_ref} and
some classic papers~\cite{classics} on cosmic topology. 

{\em Elliptical universe is unique in being the only multiply
connected FRW cosmology that preserves global isotropy and
homogeneity. This in terms of global symmetries of the universe is the
mildest possible deviation from a trivial global structure.}  As in a
simply connected FRW universe, the CMB anisotropy here is
statistically isotropic; the angular correlation function, $C(\theta)$
is simply a function of the separation implying that the angular power
spectrum ${\cal C}_\ell$ is a complete description.

\begin{figure}
\plottwoside{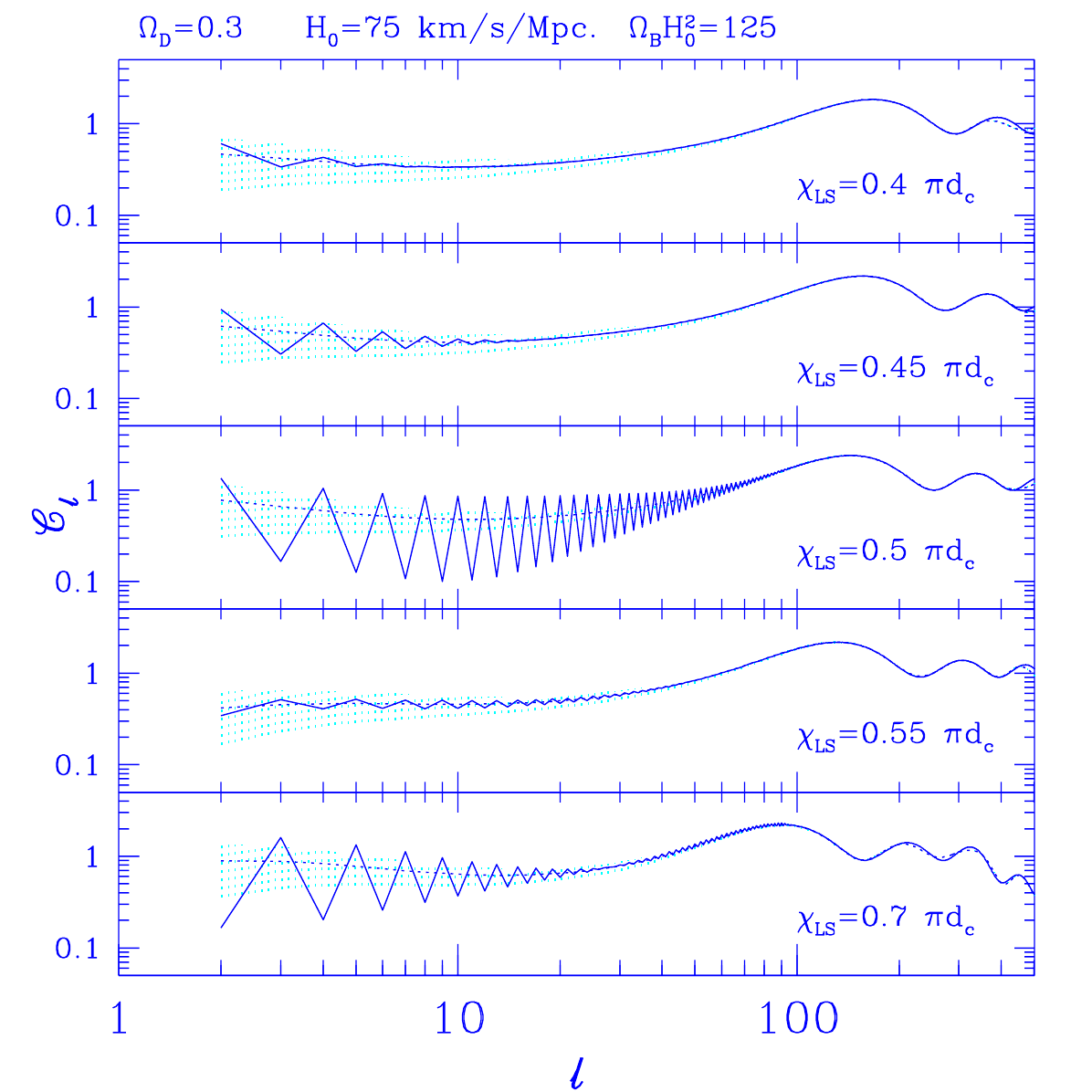}
{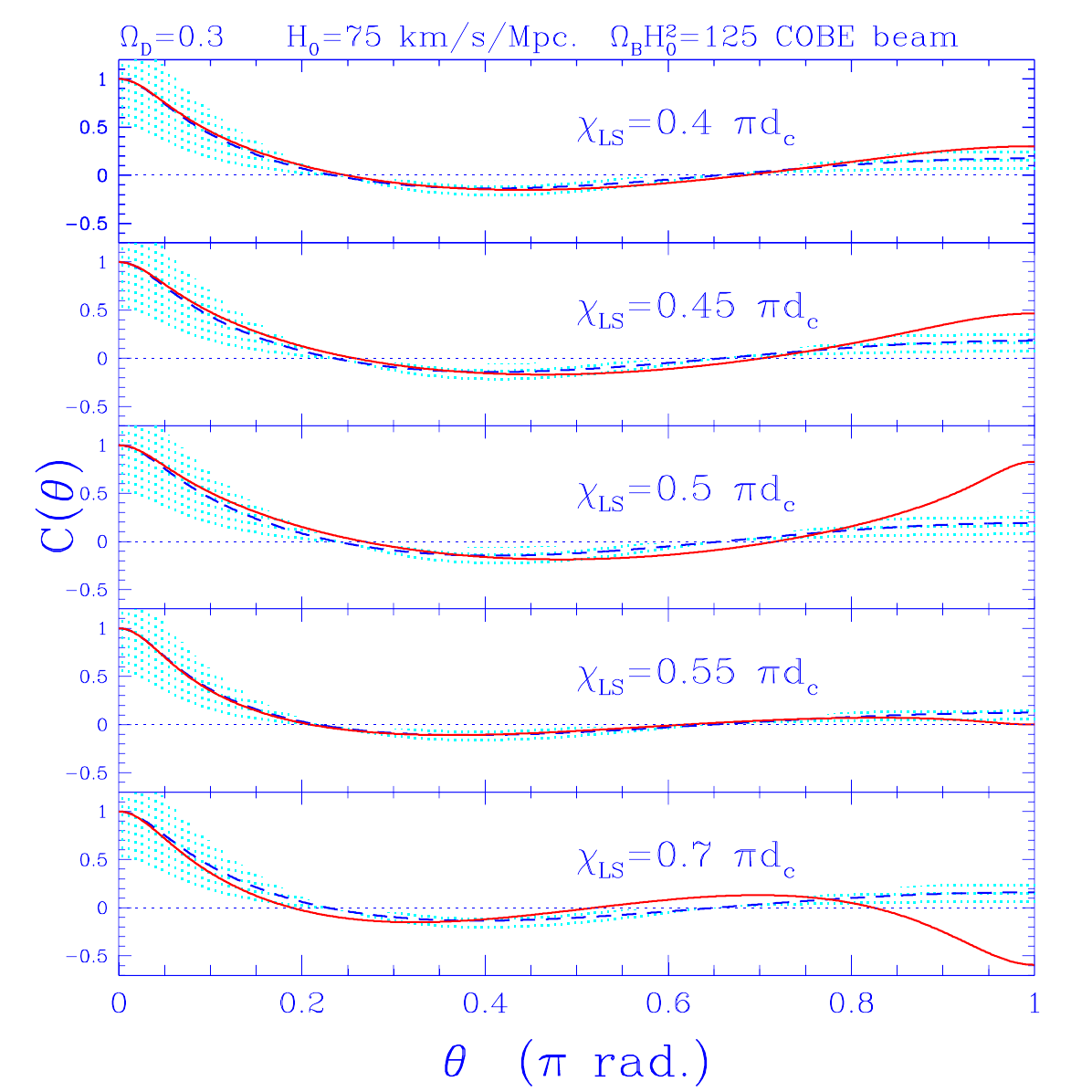}
\caption{ The figure presents a comparison of the CMB anisotropy
between the multiply connected Elliptical universe and the
corresponding simply connected (`closed') FRW universe with spherical
geometry. The density of pressure-less ``Dust'' component is fixed at
the $\Omega_D=0.3$ and the contribution of cosmological constant,
$\Omega_\Lambda$, is selected to give the stepped values of
$\chils\approx\tauH/d_c$ marked on the panels.  The {\em left panels}
show the angular power spectra, $\cl$, for the Elliptical (jagged
curve) and `closed' model (smooth) curve. The shaded band around the
closed model is an estimate of the cosmic variance. The $\cl$ are
normalised to unity at $\ell=250$ to facilitate a good visual
comparison; a COBE normalisation generally introduces an offset which
is important to bear in mind. The $\cl$'s for the Elliptical models
show an marked difference in amplitude between the odd and even values
of the harmonic, $\ell$, up to $\ell \lta 100$. For $\tauH/d_c \le
\pi/2$, the even values of $\cl$ are enhanced relative to the odd one
while the trend tends to reverse for $\tauH/d_c > \pi/2$.  This
alternating pattern of $\cl$'s is reflected strongly in the angular
correlation function, $C(\theta)$. The {\em right panels} plot
$C(\theta)/C(0)$ for the same set of models measured with a COBE
beam. The solid and dashed curves are for the Elliptical and `closed'
models, respectively; the shaded band estimates the error bar. The
ratio $C(0)/C(\pi)$ seems to be a good measure to distinguish locally
(geometrically) identical universes with and without Elliptical
topology.}
\label{fig:ellip_Cl_Cth} 
\end{figure}

Figure~\ref{fig:ellip_Cl_Cth} shows sample plots of ${\cal C}_\ell$
and $C(\theta)$ expected in Elliptical universe. The corresponding
results for the spherical universe is also plotted for comparison. The
signature of Elliptical universe is described in the caption. In this
article I present a brief explanation of the basic effect. (A more
detailed explanation will be given in a future
publication~\cite{me_ellip}.)

In the standard picture, the CMB that we observe is a Planckian
distribution of relic photons which decoupled (last scattered) from
matter at a redshift $\approx 1100$. These photons have freely
propagated over an affine distance $\chils\approx \tauH$ from this
two-sphere of last scattering (SLS) to the observer at its center.
The CMB anisotropy arises from the variations in the local properties
on the SLS such as the gravitational potential fluctuations $\to$
Surface Sachs-Wolfe effect (SSW), velocity of baryons $\to$ acoustic
peaks, \etc. and from an integral over the evolving gravitational
fluctuations encountered by the photons along its path $\to$ the
Integrated Sachs-Wolfe effect (ISW).

The CMB anisotropy signature of an Elliptical universe is on large
angular scales where the Sachs-Wolfe effect dominates the CMB
anisotropy.  The angular correlation function of the CMB anisotropy
then depends only on the spatial correlation functions on a equal-time
spatial hypersurfaces (see \cite{us_CH}). The method of images implies
that for Elliptical spaces, the correlation function, $\xi^e({\bf
x}_A, {\bf x}_B)$ between two points ${\bf x}_A\equiv(\chi_A, \hat {q}_A)$ and ${\bf x}_B\equiv(\chi_B, \hat q_B)$ can be expressed as

\begin{equation}
\xi^e({\bf x}_A, {\bf x}_B) = \xi^s({\bf x}_A,{\bf x}_B) + \xi^s({\bf
x}_A,{\bf x}_{B\prime})
\label{moi_ellip}
\end{equation}
in terms of the correlation function, $\xi^s$, on the spherical space
where ${\bf x}_{B\prime}\equiv (\pi-\chi_B, -\hat q_B)$ is the
antipode of ${\bf x}_B$. The only feature of $\xi^s$ required for this
explanation is that it falls off with increasing separation.

\begin{figure}
\plotone{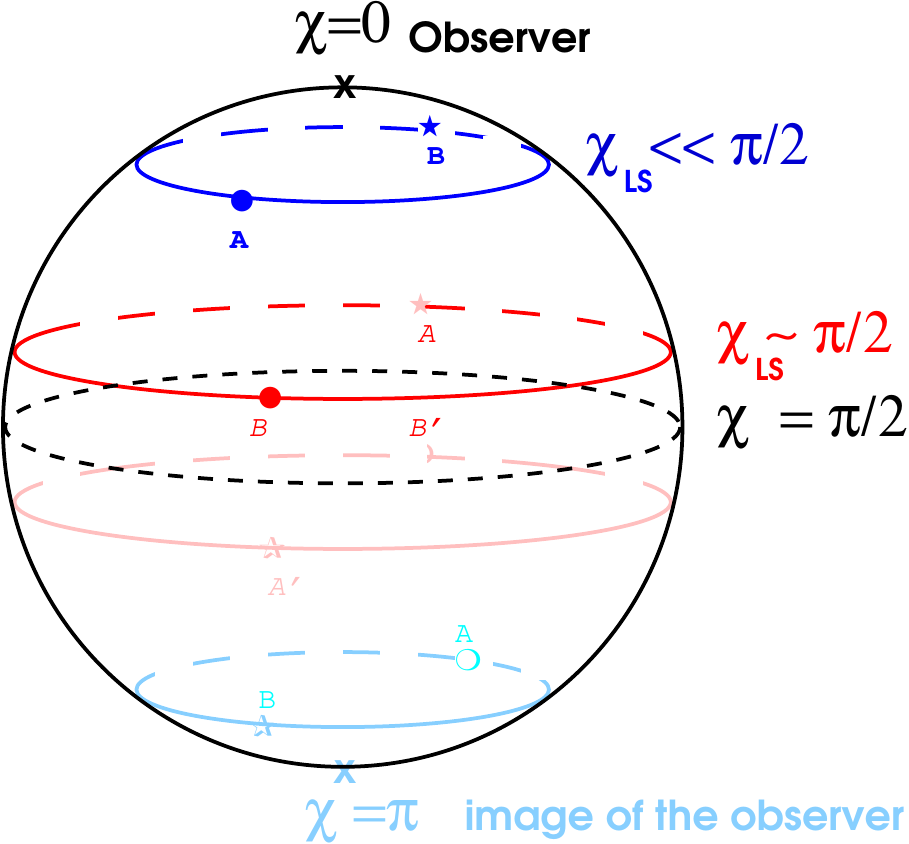}
\caption{The figure supplements the explanation based on the method of
images provided in the text for the CMB anisotropy signatures of an
Elliptical universe. The CMB in an Elliptical universe can be
interpreted as that on a simply connected spherical (`closed' FRW)
universe but with a doublet of sources. The spatial section of an
Elliptical universe is represented as a spherical manifold with
antipodal points identified. The polar angle $\chi$ measures the
radial light-travel distance from the observer, `X'; every circle of
constant $\chi$ denotes a physical two-sphere at that distance. The
spheres of last scattering (SLS) is shown for a cosmological setting
with $\tauH=\chils\ll\pi/2$ and one for which $\tauH\approx\chils \lta
\pi/2$. The points $A$ and $B$ with three coordinates $(\chils,\hat{q}
)$ and $(\chils,-\hat q)$ represent two CMB source points on the SLS
along the lines of sight in two diametrically opposite directions
$\hat q$ and $-\hat q$ in the sky.  Each source point of the CMB
anisotropy $A$ and $B$ has an image $A^\prime$ and $B^\prime$ at the
antipode of the $3$-sphere.}
\label{fig:ellip_univ} 
\end{figure}

 Consider the correlation between the CMB temperature in two
diametrically opposite directions in the sky, $\pm\hat q$, and assume
for simplicity that the surface Sachs-Wolfe effect dominates the CMB
anisotropy. The CMB correlation is then just the correlation between
the potential fluctuation on two diametrically opposite points on the
SLS.

Two cases of the above situation is illustrated in
figure~\ref{fig:ellip_univ}. For the case, when $\chils\ll\pi/2$, the
correlation $\xi^e\approx\xi^s$ between $A$ and $B$ because the second
image term in eq.(\ref{moi_ellip}) is much smaller ($A$ is far from
$B^\prime$).  Hence, CMB anisotropy in the Elliptical universe is
essentially indistinguishable from that in the corresponding closed
FRW universe.  This is to be expected since the scale of global
connectivity is $\pi/2$ in units of the curvature radius, $d_c$.

 In contrast, for the case when $\chils \sim \pi/2$, the correlation
between the point $A$ and $B$ given by eq.(\ref{moi_ellip}) is much
higher than what one get in a `closed' universe because the second
term is large owing to the proximity of $B^\prime$ to $A$. This
explains the enhanced angular correlation between diametrically
opposite directions in the sky in the Elliptical universe in the
$\chils=0.45\pi$ and $\chils=0.5\pi$ panels of
fig.~\ref{fig:ellip_Cl_Cth}. Such a correlation function translates to
a jagged ${\cal C}_\ell$ curve where the odd multipoles are suppressed
relative to the even ones. As $\chils\to\pi/2$, the pattern gets
stronger and persists up to progressively larger values of $\ell$.

In figure~\ref{fig:ellip_Cl_Cth}, the $\chils=0.7\pi$ panels show that
the CMB anisotropy spectrum and correlation functions for the
Elliptical universe also differs dramatically from its spherical
counterpart for $\chils>\pi/2$. However, the deviation is reversed:
there is now a strong anticorrelation diametrically opposite
directions in the sky, and consequently, the even harmonics in ${\cal
C}_\ell$ are now suppressed relative to its odd neighbours. This cannot
be understood solely in terms of the surface term of the Sachs-Wolfe
effect which predicts identical CMB for models with equal
$|\pi/2-\chils|$. The effect arises because the photon path from the
SLS to the observer now crosses the image of the SLS leading to a
large negative interference term between the SSW and ISW effects. This
generic correlation feature, seen in multiply connected universes
where ISW is important, was pointed out in \cite{us_CH} for compact
hyperbolic universes. There is a value of $\chils$ just beyond $\pi/2$
where the positive contribution of the pure SSW correlation to
$C(\pi)$ is cancelled by the negative contribution from the interference
term, \eg see the $\chils=0.55\pi$ panel in
fig.~\ref{fig:ellip_Cl_Cth}.

\begin{figure}
\plotone{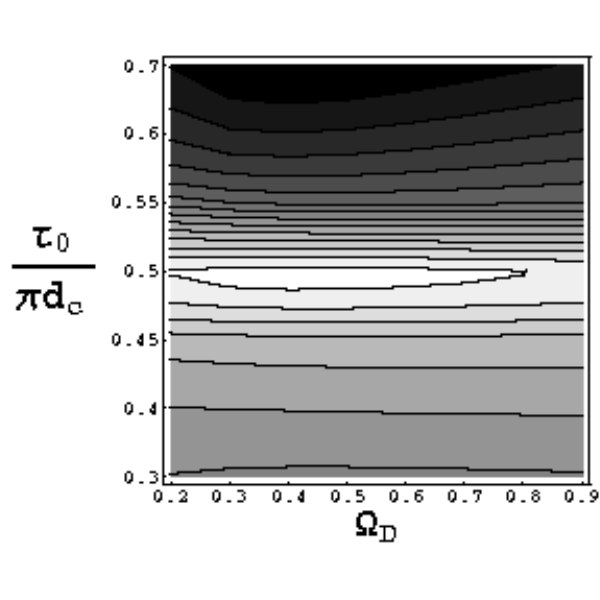}
\caption{The extreme levels in contour plot of the ratio $C(\pi)/C(0)$
for Elliptical universes in the $\Omega_D-\tauH$ parameter space are
the regions most readily probed by the CMB anisotropy data. The
correlation is computed assuming the \cobedmr beamwidth. The contours
are in steps of $0.1$, ranging from $C(\pi)/C(0)=-0.5$ to $0.8$. Since
the magnitude of ratio is $\sim 0.1-0.2$ in simply connected models, a
large region of the parameter space can be probed for Elliptical
topology using the \cobedmr data.}
\label{fig:ellip_antipod_contour} 
\end{figure}

An important signature of the Elliptical universe relative to its
spherical counterpart is the anomalously large magnitude of
$C(\pi)/C(0)$.  Fig.~\ref{fig:ellip_antipod_contour} presents a
contour plot of $C(\pi)/C(0)$, measured with a \cobedmr beam, in the
$\Omega_D- \tauH$ plane. Models in the bright (high positive values)
and dark regions (high negative values) in the contour plot predict a
potentially detectable signal in the \cobedmr data.  This overlaps
with a large portion of the preferred ($1\sigma$) region of the recent
high redshift supernova results in the $\Omega_D-\Omega_\Lambda$ plane
(see fig.~\ref{fig:omd_omv_plane}).  Roughly, only a few levels in
gray corresponding to $C(\pi)/C(0)$ between $-0.1$ to $0.4$ will be
inaccessible. This roughly corresponds to the models where either the
horizon size is too small $\tauH/d_c < 0.45\pi$, or which correspond
to the thin band just above $\tauH/d_c=\pi/2$ where two competing
effects annul the signal. This provides only a rough guide to the
results expected from the complete likelihood analysis using the full
\cobedmr data that is currently underway~\cite{me_ellip}.

It should be noted that future CMB data covering large portions of the
sky at higher resolution will be more sensitive to the signature of an
Elliptical universe. Ideally, the contrast between $C(\pi)/C(0)$ in a
Spherical and Elliptical universe can be enhanced by filtering out the
first few multipoles (this ensures the $C(\pi)/C(0)$ in spherical
model is very small) and filtering out power on $\ell\gta 100$ (since
difference between the elliptical and spherical model is negligible on
small angular scales). Since \cobedmr beam corresponds to a Gaussian
cutoff around $\ell \sim 18$ there is scope for improvement with
future `all-sky', higher resolution data.

\section{Addressing general ultra-large scale possibilities}
\label{ulss_sign}

Nontrivial topology is just one aspect of the possible nontrivial
global spatial structure of the universe. In general one could
envisage bizarre possibilities such as shown in
fig.~\ref{fig:bub_univ}.  How should one address the detectability of
a general breakdown of homogeneity and connectivity on scales just
beyond the horizon? One such general idea is the Grischuk-Zeldovich
effect which uses the small values of the detected quadrupole in the
CMB anisotropy to constrain the length scale on which universe could
have density fluctuations of order unity~\cite{gris_zel78}.

Exploiting the general connections between the geometry of manifolds
and the spectrum of eigenvalues of the Laplacian is perhaps an
interesting approach~\cite{mathlim}. The constraints from CMB
anisotropy data on the lowest eigenvalues of the Laplacian could have
a lot to say . Let me try to motivate this approach with a toy
example. Consider the bubble universe shown in fig.~\ref{fig:bub_univ}
connected to the rest of the universe by the throat. Consider the
throat to have area $A$, the volume of the bubble region to be $V$ and
the curvature radius in negatively curved throat region is greater
than $d$. Ignore the complication of the rest of the manifold and
simply assume that it is compact and has much volume greater than $V$,
and that there are no sections narrower than the throat nor regions
more negatively curved.  Given these conditions, the first eigenvalue
of the Laplacian, $k_1^2$, can be bounded using known mathematical
results to be in the range $h^2_C/4 \le k_1^2\le 4 h_C/d + 10 h_C^2$,
where $h_C = A/V$ is the Cheeger's constant.  Sharper bounds and less
restrictive conditions may be possible, but even with what we have,
one can, in principle, constrain the size of the throat. Very narrow
throat would imply small values of $h_C$ and can force the value of
$k_1^2$ to be below what is expected in a simply connected spherical
universe with curvature $d_c$. If the horizon is sufficiently large,
the presence or absence of a supercurvature mode will be discernible
in the CMB anisotropy.

\section{Conclusions}

The recent high redshift supernova results which point to the
existence of a non zero cosmological constant also have encouraging
news for the detectability of nontrivial features in the global
spatial structure of our universe using the CMB anisotropy
measurements. This general statement is borne out by this study of CMB
anisotropy in the multiply connected Elliptical universe models.

\section{Acknowledgements}
I thank Naresh Dadhich and Ajit Kembhavi for the opportunity to
contribute to this festschrift. Some of the ideas expressed here have
originated from discussions with Dick Bond and Dmitry Pogosyan during
our collaboration on the CMB anisotropy in CH universes.  The CMB
calculations were based on a `closed' universe code created by
appropriately modifying the `open' universe code in the CMBFAST
package. Part of this work was done while the author was at CITA,
Toronto. At present the author is supported by NSF grant EPS-9550487
with matching support from the state of Kansas.

\end{article}


\begin{thebibliography}{99}

\bibitem{tagore_quote} Dyson, K. K., 1992, {\sl I Won't Let you Go --selected
poems of Rabindranath Tagore}, (UBS Publishers).

\bibitem{ses_nar85} Narlikar, J. V., and Seshadri, T. R., 1985,
Astrophys. J., {\bf 288}, 43.

\bibitem{SNres} Perlmutter, S. \etal, 1999, preprint (astro-ph/9812133)
({\sl To appear in} Astrophys. J. {\bf 516}).

\bibitem{ntt} Wolf, J. A., 1994, {\it Space of Constant Curvature (5th
ed.)}, (Publish or Perish, Inc.); Vinberg, E. B., 1993, {\it Geometry
II -- Spaces of constant curvature}, (Springer-Verlag).

\bibitem{rev_ref} Lachieze-Rey, M. \& Luminet, J.-P., 1995,
Phys. Rep. {\bf 25}, 136; Starkman, G., Class. Quantum Grav. {\bf 15}, 2529,
(1998); and other articles in the same issue:  Proceedings of {\it
Topology and Cosmology}, CWRU, Cleveland, Oct. 17-19, 1997.

\bibitem{classics} Ellis, G. F. R., 1971, Gen. Rel. Grav. {\bf 2}, 7;
Sokolov, D. D., and Shvartsman, V. F., 1974,
Zh. Eksp. Theor. Fiz. {\bf 66}, 412; Gott,J. R., 1980, Mon. Not. R. Astr. Soc. {\bf 193}, 153.

\bibitem{me_ellip} Souradeep, T., {\it in preparation}.

\bibitem{us_CH} Bond, J. R., Pogosyan, D., and Souradeep, T.,
Class. Quant. Grav., {\bf 15}, 2671, (1998); {\it ibid.}, {\it to appear in
Phys. Rev. D}.

\bibitem{gris_zel78} Grishchuk, L. P. and Zeldovich, Ya. B., 1978, 
Astron. Zh., {\bf 55}, 209.

\bibitem{mathlim} Berard, P. H., 1980, {\sl Spectral Geometry: Direct and
Inverse Problems}, Lec. Notes in mathematics, {\bf 1207},
(Springer-Verlag, Berlin); Chavel, I., 1984, {\it Eigenvalues in
Riemannian geometry}, (Academic Press).

\end{thebibliography}
\end{document}